\documentclass[pra, twocolumn,showpacs,amsmath,amssymb]{revtex4}

\usepackage{graphicx}
\usepackage{dcolumn}
\usepackage{bm}

\begin{document}
\title{Precision calculation of blackbody radiation shifts for optical frequency
metrology}

\author{M. S. Safronova}
\affiliation{Department of Physics and Astronomy, University of
Delaware, Newark, DE 19716-2570, USA}

\author{M. G. Kozlov}
\affiliation{Petersburg Nuclear Physics Institute, Gatchina 188300,
Russia}

\author{Charles W. Clark}
\affiliation{Joint Quantum Institute, National Institute of Standards
and Technology and the \\University of Maryland, Gaithersburg,
Maryland, 20899-8410, USA}

\begin{abstract}

We show that three group IIIB divalent ions, B$^+$, Al$^+$, and In$^+$, have anomalously small blackbody radiation (BBR) shifts of the $ns^2 ~^1$S$_0
- nsnp~ ^3$P$^o_0$ clock transitions. The fractional BBR shifts for these ions are at least 10 times smaller than those of any other present or
proposed optical frequency standards at the same temperature, and are less than 0.3\% of the Sr clock shift. We have developed a hybrid configuration
interaction + coupled-cluster method that provides accurate treatment of correlation corrections in such ions, considers all relevant states in the
same systematic way, and yields a rigorous upper bound on the uncertainty of the final results. We reduce the BBR contribution to the fractional
frequency uncertainty of the Al$^+$ clock to $4\times10^{-19}$ at $T$=300~K.
 We also reduce the uncertainties due to this effect at room
temperature to $10^{-18}$ level for B$^+$ and In$^+$ to facilitate
further development of these systems for metrology and quantum
sensing. These uncertainties approach recent estimates of the
feasible precision of currently proposed optical atomic clocks.

\end{abstract}

 \maketitle

Development of more precise frequency standards will open ways to
more sensitive quantum-based standards for applications such as
measurements of the fundamental constants and testing of physics
postulates, inertial navigation, magnetometry,  and tracking of
deep-space probes \cite{clockbook}. Recent advances in atomic and
optical physics have led to unprecedented improvements in the
accuracy of optical frequency standards, which could lead to
redefinition of a second~\cite{si}. For example, in 2010, the most
accurate clock to date  was demonstrated, an optical clock based on
quantum logic spectroscopy of an Al$^+$ ion  \cite{Alexpt}. Its
fractional frequency uncertainty of $8.6\times10^{-18}$ is equivalent
to a shift of 1 second in 3.7 billion years.

 The definition of the second refers to a clock transition in an atom
 at  a temperature of
 absolute zero \cite{si}, whereas
all present optical atomic clocks  operate at room
 temperature (with the exception of Hg$^+$
\cite{Alexpt1}).
 Thus, the transition of a working optical clock must
 be   corrected
  for the effect of finite temperature, to which the leading contributor is the blackbody
  radiation (BBR) shift of the transition frequency.
  In fact,  the three largest systematic
 uncertainties in the Al$^+$  clock  frequency
are due to excess micromotion of the trapped ion, its secular motion,
and the BBR shift ~\cite{Alexpt}. The corresponding fractional
frequency
 uncertainties associated with these three effects were evaluated ~\cite{Alexpt} to
 be 6$\times10^{-18}$, 5$\times10^{-18}$, and 3$\times10^{-18}$.
 If the motional effects could be sufficiently suppressed by experimental techniques, the BBR
 shift, which is, in principle, calculable,
 will become the main source of uncertainty.

 Experimental measurements of the BBR shifts are sufficiently difficult that no direct
measurement has yet been reported for optical frequency standards,
and even the measurement of differential Stark shifts has only been
carried out with uncertainties  greater than 10\%. While BBR shifts
can be strongly suppressed by lowering the temperature, it is
desirable to attain the highest accuracy that is possible without
using cryogenic techniques. In this work, we have calculated the BBR
shifts in Al$^+$, B$^+$ , and In$^+$ frequency standards with 10\%
accuracy, which removes the BBR shift as a significant source of
uncertainty in the present experiments.
 Our present work calculates the  BBR shift uncertainty at constant temperature   (300~K).
Imperfect  temperature control is a source of additional experimental uncertainty.
Small BBR shifts are also favorable because they translate to
small uncertainties due to imperfect temperature control.

The BBR frequency shift of a
  clock transition can be related to the difference of the static electric-dipole
  polarizabilities between the  two clock states
\cite{Der}.
 It happens that polarizabilities of the two Al$^+$ clock states are
 nearly equal (within 2\%).
 This is a source of  difficulty in the
 calculation and the determination of its uncertainty.
 Due to this cancellation, the BBR shift in Al$^+$ frequency standard \cite{Alexpt}
 is  at least an order of magnitude smaller than that estimated for any other atomic frequency
   standard to date~\cite{IEEE,BBRexpt,BBRth,das}.
 Precise
   knowledge of the BBR shift and its uncertainty
    is essential for further improvement of
  the accuracy of the Al$^+$ optical frequency standard.
 Here we report the development of a new method of first-principles
 calculation that reduces  the relative uncertainty due
to the BBR shift at 300K  in Al$^+$  to $4\times10^{-19}$.

 In this work, we have also
    investigated other divalent group IIIB ions, B$^+$ and In$^+$,
    which have been suggested as potential optical frequency standards
\cite{b,In1,In2,In3}. No  published values of their BBR shifts exist
at the present time to the best of our knowledge. We find that B$^+$
and In$^+$ have unusually small BBR shifts due to the same type of
cancellation that we find in Al$^+$. Al$^+$ remains the species with
the smallest BBR shift yet investigated, but In$^+$ and B$^+$ are
superior in terms of the BBR shift to any other present frequency
standards.

 Unless stated otherwise, we use atomic units (a.u.) for all matrix
elements and polarizabilities throughout this paper: the numerical
values of the elementary
 charge, $e$, the reduced Planck constant, $\hbar = h/2
\pi$, and the electron mass, $m_e$, are set equal to 1. The atomic
unit for polarizability can be converted to SI units via
$\alpha/h$~[Hz/(V/m)$^2$]=2.48832$\times10^{-8}\alpha$~(a.u.). The
conversion coefficient is $4\pi \epsilon_0 a^3_0/h$ in SI units and
the Planck constant $h$ is factored out in order to provide direct
conversion into frequency units; $a_0$ is the Bohr radius and
$\epsilon_0$ is the electric constant.

The BBR frequency shift of the clock  transition  can be related to
the difference of the static electric-dipole polarizabilities between
the clock states, $\Delta\alpha_0$,  by \cite{Der}
\begin{equation}
 \delta \nu = -\frac{1}{2}(831.9~\mathrm{V/m})^2
\left( \frac{T(K)}{300} \right)^4 \Delta \alpha_0(1+\eta),
\label{ee5}
\end{equation}
where  $\eta$ is a small dynamic correction due to the frequency
distribution
 and only the electric-dipole transition part of the contribution is
considered. In this equation, $\Delta \alpha_0$ in atomic units has
to be multiplied by the numerical factor from the previous paragraph.

Precision calculations for divalent atoms require an accurate
treatment of the strong valence-valence correlations; low-order
perturbation theory does not give results of competitive accuracy for
neutral atoms or singly-charged ions.
 As a
solution to this problem, we developed an \textit{ab initio}
theoretical method within the framework of relativistic many-body
theory to accurately treat correlation corrections in divalent  atoms
\cite{CIall}.  This method combines the all-order approach currently
used in precision calculations of properties of monovalent atoms
\cite{reviewall} with the configuration-interaction (CI) approach
that is applicable for many-electron systems. Here  we report the
extension of this method to calculate ground and excited state
polarizabilities of divalent ions.

\begin{table}[ht]
\caption[]{ Comparison of calculated and experimental energies  of
 Al$^+$ in cm$^{-1}$. Column 1: level designation. Column 2:
experimental energies. The entry for the ground state $3s^2$ is its
two-electron binding energy; all excited state entries are energies
measured from the ground state.  Columns 3-5: differences of
experimental from theoretical values in CI, CI+MBPT, and CI+all-order
approximations. For example, the CI $3s^2$ energy is
$381308-4718=376591$~cm$^{-1}$.} \label{tab1}
\begin{ruledtabular}
\begin{tabular}{lrccc}
   \multicolumn{1}{c}{Level} & \multicolumn{1}{c}{Expt.} &  \multicolumn{1}{c}{CI} &
      \multicolumn{1}{c}{CI+MBPT} & \multicolumn{1}{c}{CI+All} \\[0.2pc]
\hline
$   3s^2 ~   ^1$S$_0 $   &   381308      &   4718    &   163 &   23  \\
$   3p^2 ~   ^1$D$_2 $   &   85481       &   1984    &   61  &   -19 \\
$   3s4s    ~   ^3$S$_1 $   &   91275    &   1290    &   62  &   14  \\
$   3p^2 ~   ^3$P$_0 $   &   94085       &   1499    &   34  &   7   \\
$   3p^2 ~   ^3$P$_1 $   &   94147       &   1499    &   30  &   4   \\
$   3p^2 ~   ^3$P$_2 $   &   94269       &   1498    &   23  &   -4  \\
$   3s4s    ~   ^1$S$_0 $   &   95351    &   1359    &   50  &   3   \\
$   3s3d    ~   ^3$D$_3 $   &   95549    &   1353    &   -2  &   -25 \\
$   3s3d    ~   ^3$D$_2 $   &   95551    &   1354    &   -2  &   -24 \\
$   3s3d    ~   ^3$D$_1 $   &   95551    &   1354    &   -1  &   -24 \\[0.3pc]
$   3s3p    ~   ^3$P$^o_0 $   &   37393    &   1155    &   56  &   3\\
$   3s3p    ~   ^3$P$^o_1 $   &   37454    &   1154    &   52  &   3   \\
$   3s3p    ~   ^3$P$^o_2 $   &   37578    &   1153    &   45  &   -6  \\
$   3s3p    ~   ^1$P$^o_1 $   &   59852    &   255 &   -105    &
-84
  \end{tabular}
\end{ruledtabular}
\end{table}
In the combined CI + all-order approach  used in the present work, core
 excitations are incorporated in the CI method by constructing an effective
Hamiltonian using fully converged all-order excitation coefficients
\cite{CIall}. Therefore, the core-core and core-valence sectors of
the correlation corrections for divalent systems
 are treated with the same accuracy as in the all-order approach for
monovalent atoms. Then, the CI method is used to treat valence-valence
 correlations. For divalent systems, only two-particle configuration space needs to be
considered, so the configuration space can be made numerically
complete. The valence part of the polarizability is determined  by
solving the inhomogeneous equation of perturbation theory in the
valence space, which is approximated as
\begin{equation}
(E_v - H_{\textrm{eff}})|\Psi(v,M^{\prime})\rangle = D_{\mathrm{eff},q} |\Psi_0(v,J,M)\rangle
\end{equation}
for a state  $v$ with the total angular momentum $J$ and projection
$M$ \cite{kozlov99a}. The wave function $\Psi(v,M^{\prime})$, where
$M^{\prime}=M+q$, is composed of parts that have angular momenta of
$J^{\prime}=J,J \pm 1$ from which the scalar and tensor
polarizability
 of the state $|v,J,M\rangle$ can be determined \cite{kozlov99a}. The
construction of the effective Hamiltonian  $H_\textrm{eff}$ using the
all-order approach is described
 in \cite{CIall}. The effective dipole operator
$D_{\textrm{eff}}$ includes random phase approximation (RPA) corrections. The calculations
 are carried out with a finite B-spline basis set
\cite{Bspline}, with several lower orbitals replaced by exact Dirac-Hartree-Fock (DHF) functions.

In order to establish the accuracy of our approach, we also perform the CI and CI+MBPT calculations
 carried out with the same parameters
(configuration space, basis set, number of partial waves, etc.). No core excitations are added
 in the pure divalent CI approach. In the CI+MBPT
method,   core excitations are incorporated by constructing an
effective Hamiltonian using second-order many-body perturbation
theory \cite{dzuba96b}. Comparison of the CI, CI+MBPT,  and
CI+all-order values allows us to evaluate the
 importance of the various correlation corrections,
therefore establishing the upper bound on the uncertainty of our calculations.

Table \ref{tab1} presents the comparison of the experimental energies of Al$^+$
levels with those  calculated in the CI, CI+MBPT,
  and CI+all-order approximations. The supplementary
 material contains the analogous data for B$^+$ and In$^+$ ~\cite{EPAPS}.
\begin{table}
\caption[]{ Contributions to the $3s^2 ~^1$S$_0$ and $3s3p
~^3$P$^o_0$ polarizabilities $\alpha_0$ of  Al$^+$ in a.u. Absolute
values of the corresponding reduced electric-dipole matrix elements
are listed in column labeled ``D'' in a.u. Contributions labeled
``Other'', ``Core'', and ``VC'' are described in the text. Final
polarizability values are listed in rows labeled ``Total''.}
\label{tab2}
\begin{ruledtabular}
\begin{tabular}{llcr}
 \multicolumn{1}{c}{State} &    \multicolumn{1}{c}{Contr.} & \multicolumn{1}{c}{$D$} &     \multicolumn{1}{c}{$\alpha_0$} \\
\hline
   $   3s^2 ~^1$S$_0 $   &   $  3s^2 ~^1$S$_0 - 3s3p~ ^1$P$^o_1    $   &   3.113   &   23.661  \\
            &   $ 3s^2 ~^1$S$_0 -  3s4p~ ^1$P$^o_1    $   &   0.045   &   0.003   \\
            &       Other       &       &   0.138   \\
            &       Core        &       &   0.265   \\
            &       VC      &       &   -0.019  \\
            &       Total       &       &   24.048  \\ [0.5pc]
   $   3s3p ~^3$P$^o_0 $   &   $ 3s3p ~^3$P$^o_0 - 3s4s~^ 3$S$_1    $   &   0.900  &   2.197   \\
            &   $  3s3p ~^3$P$^o_0 - 3p^2~^3$P$_0 $   &   1.836  &   8.687   \\
            &   $  3s3p ~^3$P$^o_0 - 3s3d~^3$D$_1    $   &   2.236  &   12.568  \\
            &       Other       &       &   0.836   \\
            &       Core        &       &   0.265   \\
            &       VC     &       &   -0.010  \\
            &       Total       &       &   24.543
    \end{tabular}
\end{ruledtabular}
\end{table}
 Significant improvement of the energy values is observed for Al$^+$ and In$^+$ with
 the CI+all-order method as expected due to the more complete
inclusion of the correlation corrections than in the CI and CI+MBPT
approaches. For most levels, the CI+all-order energies are within a
few cm$^{-1}$ of the experimental values for B$^+$ and Al$^+$. The
accuracy of the In$^+$ energy levels is sufficient for the purposes
of the present work: replacing our theoretical energies by the
experimental values in the dominant polarizability
 contributions changes the BBR shift by only 1\% .

The breakdown of the contributions to the $3s^2 ~^1$S$_0$ and $3s3p
~^3$P$^o_0$ polarizabilities
 $\alpha$ of  Al$^+$ is given in
Table~\ref{tab2}. The supplementary
 material contains the analogous data for B$^+$ and In$^+$ ~\cite{EPAPS}.
While we do not use the sum-over-state approach in the calculations of the polarizabilities,
 it is useful to establish which terms give the dominant
contributions.  We evaluate several dominant contributions to
polarizabilities
 by combining our values of the E1 matrix elements and energies as
${2D^2}/{3\Delta E}$ according to the sum-over-states
formula~\cite{JPBreview}
 with $J=0$. We find that
a single transition, $ ns^2 ~^1$S$_0 -  nsnp~^1$P$^o_1$, contributes
92.9\%, 99.4\%, and 98.7\% to the valence ground state polarizability
for B$^+$, Al$^+$, and In$^+$, respectively. Three transitions, $nsnp
~^3$P$^o_0 - np^2~^3$P$_1 $, $nsnp ~^3P_0 - ns(n+1)s~^3$S$_1$, and
$nsnp~^3$P$^o_0 - ns(n+1)d~^3D_1$ contribute 70.8\%, 98.7\%, and
92.8\% to the $^3$P$^o_0$ polarizability for B$^+$, Al$^+$, and
In$^+$, respectively. Therefore, both Al$^+$ and In$^+$
polarizabilities could be calculated more precisely if experimental
values of the dipole matrix elements were known to high precision. We
subtract the values of the terms listed separately in
Table~\ref{tab2} from our total valence polarizability values to
obtain the remaining contributions that are listed in the rows
labeled ``Other". Our dominant contributions for Al$^+$ are in
excellent agreement with CI calculations with a semi-empirical core
potential (CICP) \cite{BBRth}.

The ionic core polarizability and VC term  that corrects it  for
 the presence of the valence electrons are listed in rows labeled
``Core" and ``VC''. We note that the ionic core contribution is the
same for both
 clock states and so it does not contribute to the BBR shift. On the other hand, the VC
contribution is different for the two clock states. It is negligible for B$^+$. It is the largest for the $^3$P$^o_0$ polarizability of In$^+$ to
which it contributes only 0.5\%. However, its contribution to the BBR shift is much larger, 1.8\% and 5\% in Al$^+$ and In$^+$, respectively, owing
to the large degree of cancellation between $^1$S$_0$ and $^3$P$^o_0$ polarizabilities. We estimate the dominant uncertainty in this term as the
difference of the DHF and RPA values, and assume that all other uncertainties do not exceed this dominant uncertainty. Adding these two uncertainties
in quadrature, we estimate that VC term leads to the 0.6\% and 2\% uncertainties in the BBR shifts for Al$^+$ and In$^+$.
\begin{table}
\caption[]{ The values of the $ns^2 ~^1$S$_0$ and $nsnp ~^3$P$^o_0$
polarizabilities $\alpha_0$ in
 B$^+$, Al$^+$, and In$^+$ calculated in CI, CI+MBPT,
and CI+all-order approximations in a.u.  CI+all-order values are
taken as final. } \label{tab3}
\begin{ruledtabular}
\begin{tabular}{llrrr}
\multicolumn{1}{c}{Ion} &  \multicolumn{1}{c}{} &    \multicolumn{1}{c}{CI}
& \multicolumn{1}{c}{CI+MBPT} &     \multicolumn{1}{c}{CI+all} \\
\hline
B$^+$  &   $   \alpha_0(2s^2~ ^1$S$_0) $   &   9.575   &   9.613   &   9.624   \\
    &   $   \alpha_0(2s2p~ ^3$P$^o_0)  $   &   7.779   &   7.769   &   7.772   \\
    &   $   \Delta \alpha_0  $   &   -1.796  &   -1.844  &   -1.851  \\[0.5pc]
Al$^+$ &   $   \alpha_0(3s^2~ ^1$S$_0) $   &   24.405  &   24.030  &   24.048  \\
    &      $   \alpha_0(3s3p~ ^3$P$^o_0)  $   &   24.874  &   24.523  &   24.543  \\
    &   $   \Delta \alpha_0  $   &   0.469   &   0.493   &   0.495   \\[0.5pc]
In$^+$ &   $   \alpha_0(5s^2~ ^1$S$_0) $   &   26.27   &   23.83   &   24.01   \\
    &   $   \alpha_0(5s5p~ ^3$P$^o_0)  $   &   28.60   &   25.87   &   26.02   \\
    &   $   \Delta \alpha_0  $   &   2.33    &   2.04    &   2.01    \\
    \end{tabular}
\end{ruledtabular}
\end{table}
\begin{table*}[ht]
\caption[]{ BBR shifts at $T=300K$ in   B$^+$, Al$^+$, and In$^+$.
$\Delta\alpha_0$ is given  in a.u.; clock frequencies $   \nu_0$ and
the BBR shifts $\Delta\nu_{\textrm{BBR}}$  are given in Hz.
Uncertainties in the values of $\Delta\nu_{\textrm{BBR}}/\nu_0$ are
given in column labeled ``Uncertainty''.} \label{tab4}
\begin{ruledtabular}
\begin{tabular}{lrrrrrrr}
\multicolumn{1}{c}{Ion} &\multicolumn{1}{c}{$\Delta \alpha_0$}
&\multicolumn{1}{c}{$\eta(^1S_0)$} &\multicolumn{1}{c}{$\eta(^3P_0)$}
&\multicolumn{1}{c}{$\Delta\nu_{\textrm{BBR}}$ (Hz) }
&\multicolumn{1}{c}{$   \nu_0$ (Hz)   } &\multicolumn{1}{c}{$
|\Delta\nu_{\textrm{BBR}}/\nu_0| $}
&\multicolumn{1}{c}{Uncertainty }      \\
\hline
B$^+$     & -1.851         &   0.00014 &   0.00014 &   0.0159(16)  &   $   1.119\times10^{15}  $   &   $   1.42\times10^{-17}  $   &   $   1\times10^{-18} $   \\
Al$^+$& 0.495              &   0.00022 &   0.00024 &   -0.00426(43)&   $   1.121\times10^{15}  $   &   $   3.8\times10^{-18}  $   &   $   4\times10^{-19} $   \\
In$^+$ &2.01               &   0.00018 &   0.00019 &   -0.0173(17) &   $   1.267\times10^{15}  $   &   $   1.36\times10^{-17} $   &   $   1\times10^{-18} $   \\
    \end{tabular}
\end{ruledtabular}
\end{table*}

There are three other major sources of uncertainties in our
calculations of the BBR shift. One is the omission of the Breit
interaction in our calculations.  We have estimated the main part of
the Breit correction  by incorporating  the one-body part of the
Breit interaction into the basis set orbitals on the same footing
with Coulomb interaction. All calculations were then repeated with
the modified basis set. The change in the Al$^+$ BBR shift was found
to be only 1.4\%.
 The two other main sources of uncertainty are
incompleteness of treatment of core excitations via the effective
Hamiltonian technique
 described above (for example, our all-order method is
restricted to single and double excitations), and limiting the
treatment of the effective dipole
 operator $D_{\textrm{eff}}$ to the RPA method \cite{kozlov99a}. The second issue is
unlikely to cause large errors as RPA is expected to be the dominant
contribution for these E1 matrix elements. Moreover, we have verified
that our \textit{ab initio} CI+all-order method reproduces the
recommended values \cite{Der} of clock state polarizabilities of Mg,
Ca, and Sr ~\cite{sr}.
 We investigate the uncertainty due to the inclusion of the core excitations by comparing
  the difference $\Delta\alpha_0$ calculated in the CI, CI+MBPT, CI+all-order approximations.
  These results are summarized  in Table~\ref{tab3}.
   We find that the entire contribution of
  core excitations to the BBR shift, estimated as the difference of the $\Delta\alpha_0$
  CI+all-order and CI values is only 3\%, 5\%, and 16\%
for B$^+$, Al$^+$, and In$^+$, respectively. The difference between CI+MBPT and CI+all-order
 values is 0.4\% for  B$^+$ and Al$^+$, and 1.7\% for
In$^+$. Therefore, we place an upper bound on the uncertainty of our BBR values at
 10\% for all three cases.

Our final results are summarized in Table~\ref{tab4}, where we list the polarizability
 difference $\Delta\alpha_0$, dynamic corrections $\eta,$ BBR
shift at $T=300~$K, $^1$S$_0-^3$P$^o_0$ clock frequencies $\nu_0$,
 relative BBR shift $\Delta\nu_{\textrm{BBR}}/\nu_0$, and the uncertainty in the relative
 BBR shift for B$^+$,
Al$^+$, and In$^+$.
 Dynamic corrections are very small for both states and nearly
cancel each other. Their contributions to the BBR shift are negligible for all three ions. We estimated that contribution to the BBR shift due to the
$^3$P$^o_0 - ^3$P$^o_1$ M1 transition is below $10^{-5}$~Hz and is negligible at the
 present level of accuracy.

 Our BBR shift value in Al$^+$, $\Delta\nu_{\textrm{BBR}}=-0.00426(43)$~Hz, is in
 agreement with CICP value of Ref.~\cite{BBRth}
 $\Delta\nu_{\textrm{BBR}}=-0.0042(32)$ and is consistent with experimental
  measurement $\Delta\nu_{\textrm{BBR}}=-0.008(3)$~Hz
from Ref.~\cite{BBRexpt}. The values of $\eta$ for Al$^+$ are in agreement with \cite{BBRth}.

Our value for the Al$^+$ BBR shift is in agreement with that of the
coupled-cluster calculation $\Delta\nu_{\textrm{BBR}}=-0.0041(7)$~Hz
\cite{das}. Although the uncertainty in that calculation was
estimated at 17\%, the individual state polarizabilities, and more
significantly, their difference $\Delta \alpha_0$, varied
considerably with choice of basis set. Specifically, the
single-double coupled cluster (CCSD) values of $\Delta \alpha_0$ with
increasing basis sets are reported to be 0.165, 0.058, 0.897, 0.427,
and 0.406 (in a.u.). Although the last two numbers are close, this
sequence in itself does not demonstrate convergence (twice the
difference  in the last two values was taken to be the
  basis set error). Heavy dependence of the polarizability on the choice of basis set
 is a well-known problem  in coupled-cluster methods (see \cite{JPBreview} and
 references
   therein). This issue was exacerbated in \cite{das} by use of different methods
   for the lower and upper clock states. In our work, on the other hand, large basis
   sets were used in all all-order/pertubation theory
  calculations (385 orbitals with $l < 6$),  and the CI space was saturated until the error was
  negligible at the present level of accuracy. The same approach and
  basis sets were used in all our calculations.
In addition to the polarizabilities, the only other numerical result
reported in Ref.~\cite{das}  is the $3s^2 - 3s3p~^3$P$^o_0$ clock
transition energy. Its reported value of 37326(95)~cm$^{-1}$ differs
from the experimental value 37393~cm$^{-1}$ by 67~cm$^{-1}$. Our
CI+all-order value 37390~cm$^{-1}$ agrees  with experiment to
3~cm$^{-1}$, and even our CI+MBPT value agrees to 56~cm$^{-1}$ (see
line $3s3p~^3$P$^o_0$ in Table~\ref{tab1}). Most of our other energy
levels also agree with experiment to a few cm$^{-1}$.

We also calculated frequency-dependent polarizabilities  of clock
 states at 1126~nm using the same approach; our values are
$\alpha_0(\omega)[^1S_0]=24.58$~a.u. and $\alpha_0(\omega)[^3P_0]=25.13$~a.u.
The resulting frequency-dependent polarizability difference
$\Delta\alpha_0(\omega)=0.549(55)$~a.u is in agreement with theoretical CICP value 0.54(41)~a.u.
 \cite{BBRth} and is $1.6\sigma$ from the
experimental value 1.08(34)a.u.~\cite{BBRexpt}.

In summary, our calculations of the BBR shifts reduce the
uncertainties in the fractional frequency shift at room temperature
to $10^{-18}$ in B$^+$  and In$^+$ and to
 $4\times10^{-19}$  in Al$^+$.
 These uncertainties approach recent estimates of the feasible
precision of currently proposed optical atomic clocks
~\cite{Alexpt1}. This work introduces a novel computational approach
that can be used for a variety of problems of importance to atomic,
nuclear, and high-energy physics, as well as quantum chemistry (study
of parity violation, searches for electron dipole moment, study of
degenerate gases, determination of nuclear magnetic moments, search
for variation of fundamental constants, etc.)

 This research was
performed under the sponsorship of the US Department of Commerce, National Institute of Standards and Technology, and was supported by the National
Science Foundation under Physics Frontiers Center Grant PHY-0822671 and by the Office of Naval Research. The work of MGK was supported in part by
RFBR grant \#11-02-00943.

\end{document}